\RequirePackage{fix-cm}
\documentclass{svjour3}                     
\smartqed  
\usepackage{graphicx}
\usepackage{mathptmx}      

\usepackage{amsmath}
\usepackage{amsfonts}
\usepackage{amssymb}
\usepackage{lmodern}
\usepackage{xfrac}
\usepackage{graphicx}
\graphicspath{ {./figures/} }
\usepackage{epsfig}
\usepackage{epstopdf}
\usepackage{textcomp}
\usepackage{multirow}
\usepackage{setspace}
\usepackage{fullpage}
\usepackage{float}
\usepackage{caption}
\usepackage{subcaption}
\usepackage{siunitx}
\usepackage[sort&compress,numbers]{natbib}
\usepackage{xcolor}
\journalname{JOM}
\begin{document}

\title{Single Track Melt Pool Measurements and Microstructures in Inconel 625
}


\author{Supriyo Ghosh     \and
        Li Ma	\and
        Lyle~E.~Levine	\and
        Richard~E.~Ricker \and
        Mark~R.~Stoudt \and
        Jarred~C.~Heigel \and
        Jonathan~E.~Guyer
}

\institute{Materials Science and Engineering Division, National Institute of Standards and Technology, 100 Bureau Drive, Gaithersburg, MD 20899, USA \\
}

\date{Received: date / Accepted: date}

\maketitle

\begin{abstract}
We use single track laser melting experiments and simulations on Inconel 625 to estimate the dimensions and microstructures of the resulting melt pools. Our work is based on a design-of-experiments approach which uses multiple laser power and scan speed combinations. Single track experiments generate melt pools of certain dimensions. These dimensions reasonably agree with our finite element calculations. Phase-field simulations predict the size and segregation of the cellular microstructures that form along the melt pool boundaries for the solidification conditions that change as a function of melt pool dimensions.
\end{abstract}

\section{Introduction}
Additive manufacturing (AM) technologies are rapidly being used to fabricate near-full density and high-value metallic parts for aerospace, medical, and defense applications~\cite{Murr2012,King2015,Francois2017}. When a laser source with a given power ($P$) and power distribution scans at a programmed speed ($V$) across an alloy substrate or powder, a melt pool of certain dimensions (length $l$, width $w$ and depth $h$) results. These dimensions are important, as they determine the density of the parts through subsequent solidification processes which affect the material properties. 

Measurements of melt pool dimensions have been reported for various combinations of $P$ and $V$ using single track experiments on Ni alloys~\cite{Heigel2017,Heigel2017b,Montgomery2015}, stainless steels~\cite{Childs2005}, and Ti alloys~\cite{Gong2014}. High values of $P$ and low values of $V$ resulted in large melt pools in the above experiments. A thermographic camera was used to estimate the thermal history in the material, from which the melt pool dimensions were determined~\cite{Heigel2017,Brandon2015}. Although the melt pool shape was similar in both powder and substrate cases, the powder case was complicated by surface roughness of the unmolten powder particles and by width and height variations and pore formations through keyholing events~\cite{Childs2005,Gong2014}. The molten pools that form during experiments~\cite{Heigel2017,Heigel2017b} on an Inconel 625 (IN625) surface become fully developed and reach a steady state for a single line laser scan after the first \SI{1}{\mm} to \SI{2}{\mm}. However, when the experiments were performed with multiple laser scans, the melt pool length deviated from the steady state as the scan line count increased~\cite{Heigel2017b}. Therefore, a single track laser scan on a bare IN625 substrate is used for validation simulations.

The interaction between the laser beam and the irradiated material can be very complex~\cite{King2015,King2014,Khairallah2016}, leading to rapid vaporization of the melt surface under the beam, which results in a recoil pressure coupled with Marangoni and other hydrodynamic effects within the melt pool~\cite{King2015,King2014,Khairallah2016}, determining the steady state melt pool dimensions in real systems. Modeling of such a multi-physics phenomenon is extremely difficult within a feasible computational time and resources. Therefore, laser melting processes with a relatively low power beam are often modeled using finite element methods with simplified assumptions of the melt pool dynamics in order to simplify calculations without any qualitative changes in the resultant melt pool shapes~\cite{Montgomery2015,Guo2000,Gong2015}.

The shape of a melt pool is important from a solidification perspective. Solidification begins in the trailing edge of the melt pool, and the temperature gradient $G$ and the solid-liquid interface growth rate $R$ change as a function of the melt pool dimensions. A rapid solidification condition~\cite{kurzbook,Rappazbook} is reported~\cite{Suresh2016,Trevor2017} to exist along the melt pool boundary and the resultant local solidification features, such as the grain size and the solute distribution increasingly deviate from local equilibrium with increasing $G$ and $R$ combinations~\cite{Gong2015,supriyo2017}. There have been no studies to understand the single track melting, solidification, and microstructure evolution phenomena in IN625. Design-of-experiments~\cite{Heigel2017,Gong2014,Li2015} based modeling of melt pool dimensions and microstructures can potentially be used for rapid optimization of $P$ and $V$ combinations to produce high-fidelity parts, since it can save significant time and cost when compared with conducting numerous physical experiments. In what follows, we present the laser single track experiments for melt pool shapes, finite element simulations for melt pool shapes and temperatures, and phase-field simulations for melt pool solidification microstructures.

\section{Melt pool shape}
\subsection{Single track experiments}
Well separated, single laser tracks on a solid IN625 plate were produced using a commercial laser powder bed fusion (LPBF) machine with a Nd:YAG laser. The IN625 plate had approximate dimensions of 25.4 mm $\times$ 25.4 mm $\times$ 3.2 mm. The plate was polished to a controlled surface finish (400 grit) and annealed at 870~$^\circ$C for 1 hour in vacuum to relieve residual stresses. Seven laser tracks were produced, each 4 mm long, with different combinations of $P$ and $V$. The dynamic lengths of the melt pools were measured \emph{in situ} using thermography, as reported and described previously~\cite{Heigel2017}. The widths of the laser tracks were measured \emph{ex situ} using a confocal laser scanning microscope (CLSM). The laser tracks were cross sectioned at the center positions (after the melt pool reaches a steady state length) and polished prior to chemical etching with aqua regia. These cross sections were then imaged using the same CLSM. 

\begin{figure*}[ht]
\begin{center}
\includegraphics[scale=0.7]{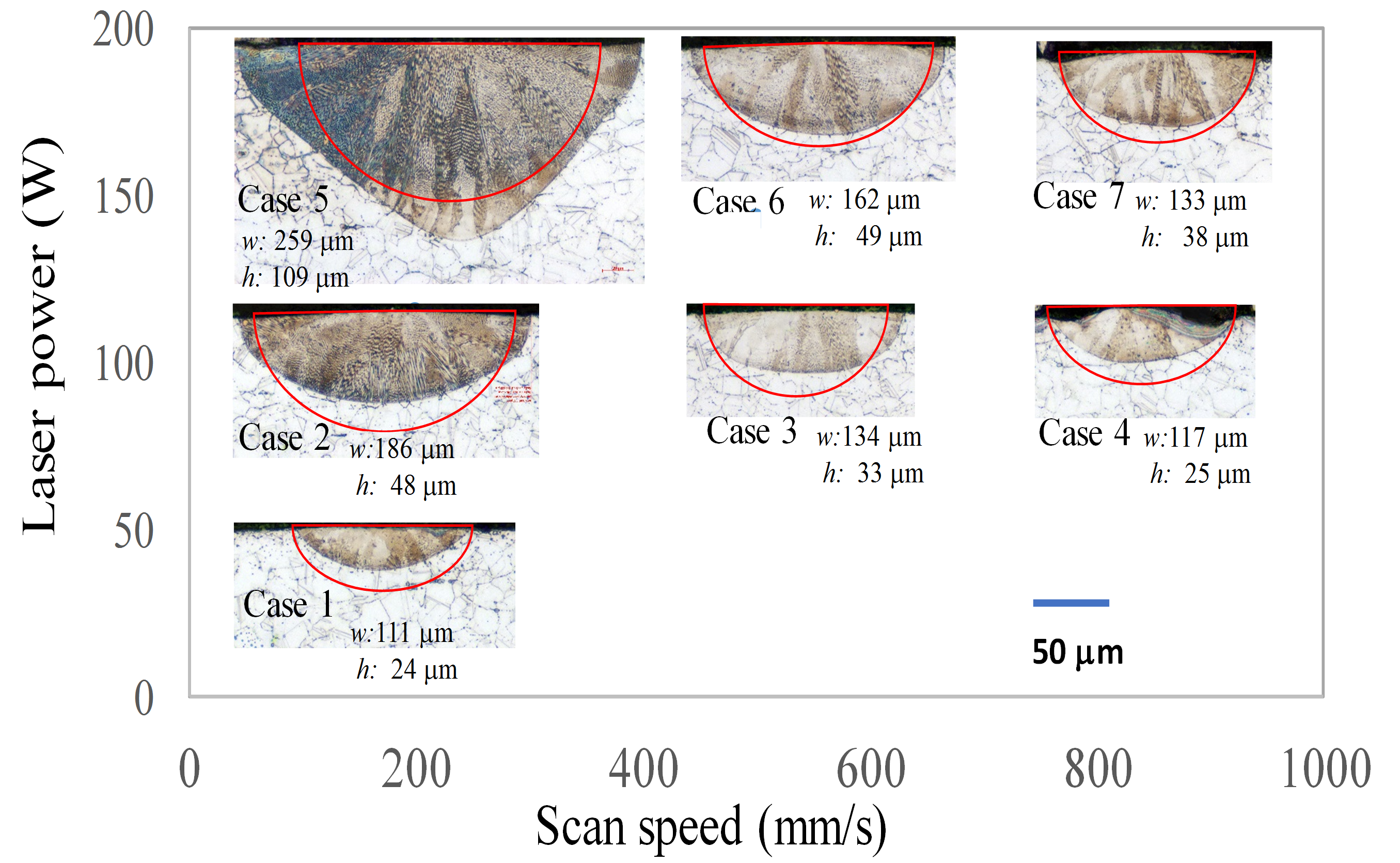}
\caption{The melt pool dimensions ($w$ and $h$) that form during experiments are measured from transverse cross sections of the laser tracks. The results are presented for seven various combinations of $P$ (= 49 W, 122 W and 195 W) and $V$ (= 200 mm s$^{-1}$, 500 mm s$^{-1}$ and 800 mm s$^{-1}$). Both $w$ and $h$ increase with increasing $P$ for a constant $V$, while they both decrease with increasing $V$ for a constant $P$. The corresponding melt pool boundary from finite element simulations is approximated by the solidus temperature isotherm (shown in red) and overlaid on the experimental images for representaion.}\label{fig_experiment}
\vspace{-8mm}
\end{center}
\end{figure*}

Figure~\ref{fig_experiment} shows the CLSM images of the laser track cross sections as a function of $P$ and $V$. Clearly, the cross section geometry is highly sensitive to both $P$ and $V$. Except for the laser track formed using 195 W and 200 mm s$^{-1}$ (Case 5), the cross sections exhibit straightforward systematic changes with respect to both $P$ and $V$. The Case 5 cross section has a different shape indicative of the onset of keyholing. Another case worth mentioning is Case 7, with 195 W and 800 mm s$^{-1}$. These are the nominal settings used for building solid components from IN625 powder for the LPBF machine used in this study. The measured melt pool width ($w$), and melt pool depth ($h$) are listed below each image. The standard deviation measurement uncertainties for the $w$ and $h$ measurements are approximately \SI{1}{\micro\meter}. The uncertainties in the melt pool length are described in~\cite{Heigel2017} and are shown graphically in Fig.~\ref{fig_length}. 

\subsection{Finite element simulations}
We perform heat transfer based finite element simulations and validate the resulting melt pool dimensions and temperature distributions using the above experiments~\cite{Heigel2017,Heigel2017b}. Similar to the experiments~\cite{Heigel2017,Heigel2017b}, a single track laser scan on a solid IN625 substrate of 25.4 mm $\times$ 25.4 mm $\times$ 3.2 mm at initial uniform temperature of 293 K is modeled. The laser beam is modeled to have a Gaussian power distribution with a $1/e^2$ radius of \SI{70}{\micro\meter}~\cite{Dayal2014}, which scans at a velocity $V$ for a length of 4 mm. Seven combinations of $P$ (49 W to 195 W) and $V$ (200 m s$^{-1}$ to 800 m s$^{-1}$) are designed for simulations, which were based on the LPBF measurements~\cite{Heigel2017,Heigel2017b}. The finite element thermal model and surface laser heat input are based on our previous works~\cite{Trevor2017,Li2015}. We consider the bulk material properties from Refs.~\cite{Trevor2017,Li2015} with no powder material or melt-state change considered. As a first approximation, the thermal conductivity is treated as a function of temperature only~\cite{Trevor2017}. The temperature-dependent bulk material density, latent heat and specific heat are calculated using thermodynamic calculations~\cite{Trevor2017}. For more details, please refer to~\cite{Trevor2017,Li2015}.

\begin{figure}[ht]
\begin{center}
\includegraphics[scale=0.5]{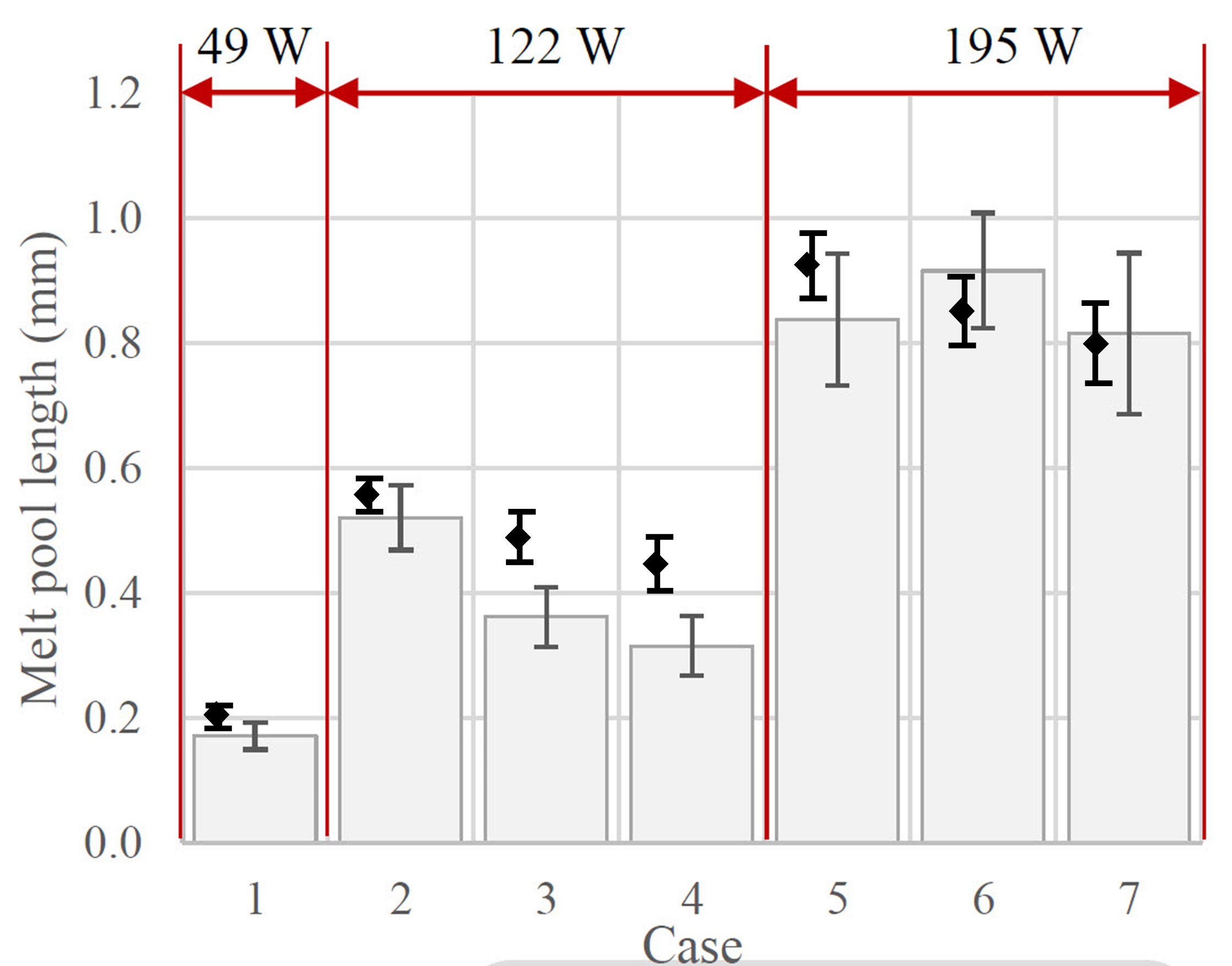}
\caption{The comparison of the melt pool length between experiments (columns) and simulations (diamonds) for the seven cases is presented. The experimental uncertainties (error bar) reflect the standard deviation variability of the melt pool length within steady state conditions~\cite{Heigel2017}. The uncertainties in the simulated melt pool lengths reflect the solidus or the liquidus transition for IN625. For most cases, simulated melt pool lengths are within experimental uncertainties. Note that the figure is our original artwork and only the data were used from Ref.~\cite{Heigel2017}.}\label{fig_length}
\vspace{-8mm}
\end{center}
\end{figure}

Although not shown here, the temperature is maximum at the top surface of the melt pool and varies along the melt pool length, width and height. The leading edge (in the direction of length) of the melt pool close to the laser is wide while the trailing edge of the melt pool is narrow as the material cools with increasing distance from the laser, consistent with the experiments~\cite{Heigel2017,Heigel2017b}. The periphery of such a typical melt pool shape is given by the solidus temperature (1587 K) isotherm for IN625, from which we extract the melt pool dimensions (Fig.~\ref{fig_experiment}). The melt pool length stabilizes typically after 2 mm from the start of the laser scan in our simulations, reaching a steady state value. Figure~\ref{fig_length} presents the simulated melt pool length measured between \SI{195}{\micro\meter} and \SI{910}{\micro\meter} at steady state, which compares reasonably with the length values from the experiments~\cite{Heigel2017}. The melt pool length increases as the laser power increases and decreases similarly as the beam speed increases, consistent with the experimental observations~\cite{Heigel2017}. The melt pool length is only 5\% smaller compared to the experimental value with 195 W and 800 mm s$^{-1}$ (Case 7), and well within the experimental uncertainties.

Next, we present the width and height measurements of the melt pool cross section having maximum depth in the build direction (Fig.~\ref{fig_experiment}). Consistent with our experiments, both $w$ and $h$ increase with increasing $P$ for a constant $V$ and decrease with increasing $V$ for a constant $P$; $w$ varies between \SI{120}{\micro\meter} and \SI{180}{\micro\meter}, and $h$ varies between \SI{38}{\micro\meter} and \SI{88}{\micro\meter}. The calculated $w$ is within 2\% of the experimental value with $V$ = 800 mm s$^{-1}$. Combining the measurements of $l$ and $w$, on average, the simulated $l/w$ are within 6\% of the experimental measurements.

The model systematically overpredicts the melt pool depth for all the Cases (except Case 5), which is presumably due to the use of a large laser absorptivity ($\eta$ = 0.5) in our simulations~\cite{King2015,Khairallah2016}. The model underpredicts the width and depth and produces an incorrect melt pool shape for Case 5, which can be attributed to complex melt pool dynamics, such as the surface tension and recoil pressure, which we do not model as a first approximation. Case 5 signifies the onset of keyholing in experiments~\cite{Heigel2017,Heigel2017b} in which a topological depression is formed owing to the vaporization of the material under high energy input and the melt pool becomes $V$-shaped.

\section{Melt pool solidification}
\subsection{Experimental microstructures}
\begin{figure*}[ht]
\begin{center}
\includegraphics[scale=0.65]{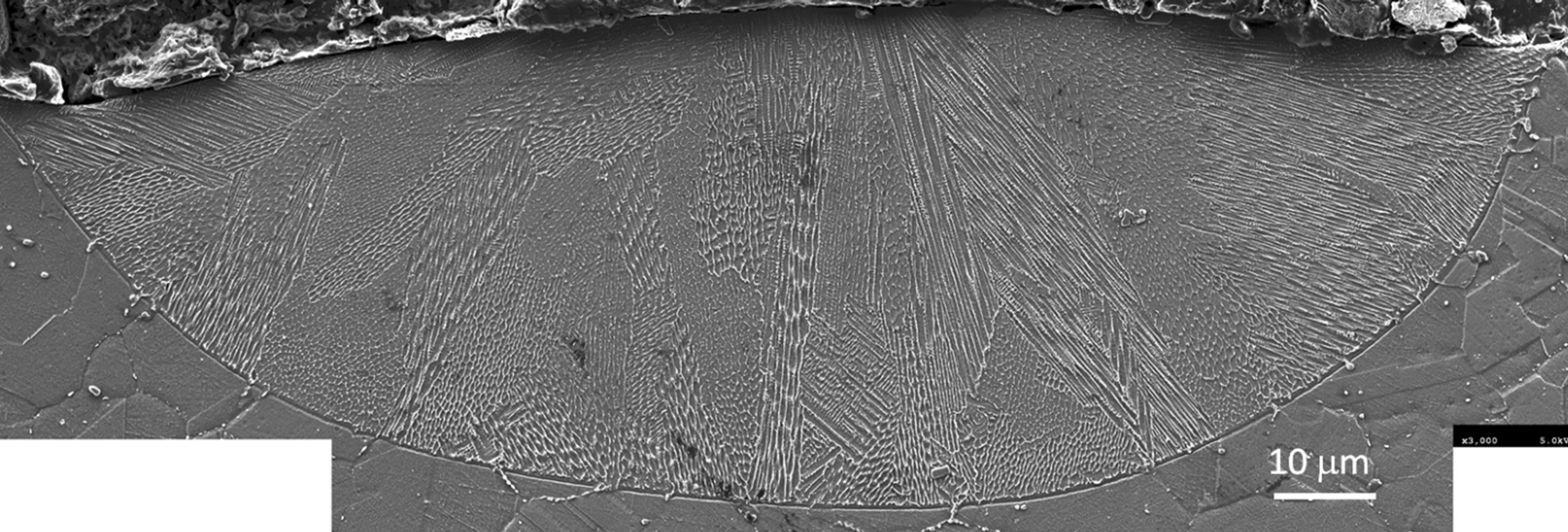}
\caption{The SEM image of the transverse cross section of the melt pool for Case 7 (Fig.~\ref{fig_experiment}) is shown. On average, the cellular dendritic microstructural features evolve roughly perpendicular to the melt pool boundary in the direction of temperature gradient.}\label{fig_microstructure}
\end{center}
\end{figure*}

A scanning electron microscopy (SEM) analysis with standard metallographic techniques~\cite{Vander1999} was used to reveal the microstructure of the solidified melt pool for 195 W and 800 mm s$^{-1}$ (Case 7). The initial solid-liquid interface can be easily distinguished in Fig.~\ref{fig_microstructure}, which evolves primarily into a cellular dendritic microstructure at steady state. On average, the microstructure is coarse close to the bottom and fine close to the rear of the solidified melt pool interface~\cite{Murr2012,Suresh2016,supriyo2017}. The individual cellular grains with distinct misorientations predominantly grow throughout the melt pool, which are often observed to solidify in the same misorientation during multi-track solidification~\cite{Murr2012,Suresh2016}. The average spacing between the primary dendrites varies between $\approx$~\SI{0.2}{\micro\meter} and \SI{1.0}{\micro\meter}, and the average grain size consisting of dendrites having the same orientation ranges between $\approx$~\SI{10}{\micro\meter} and \SI{50}{\micro\meter}. Although not shown here, one could expect from an electron backscatter diffraction analysis to show a strong, fine-scale microsegregation of Nb, Mo and others in between the dendrites~\cite{Zhang2017_nist,Eric2017}. The dendrite sizes, orientations, morphologies, and microsegregation are different at different locations within the melt pool. This is primarily due to different positions and orientations of the initial nuclei combined with different thermal gradients and solidification velocities along the melt pool boundary. \emph{In situ} measurements of these solidification conditions are difficult due to the high temperature and small volume of the melt pools, and finite element simulations are used instead.
\vspace{-1cm}
\subsection{Solidification parameters}
Solidification parameters are extracted from the simulated temperature and velocity fields along the melt pool dimensions. The trailing edge of the moving melt pool is the solidification front where the microstructural evolution takes place. The resulting complex solid-liquid interface morphologies such as the cells and dendrites can be characterized locally within the trailing edge of the melt pool by the average temperature gradient $G$ = $\frac{1}{N}\sum \nolimits_{i = 1}^{N}~|\vec{\nabla}T_i(l, w, h)|$ and the average solidification rate $R = \frac{1}{N} \sum \nolimits_{i = 1}^{N} V \cos \alpha_i(l, w, h)$. $N$ is the number of elements present between the bottom and the rear of the melt pool along the solidification front, $T_i$ is the temperature, and $\alpha_i$ is the solidification angle at element $i$~\cite{supriyo2017}. Both $G$ and $R$ vary as a function of the melt pool dimensions. We note that both $G$ and $R$ increase as the melt pool becomes smaller with increasing $V$ (Fig.~\ref{fig_phase-field}). On the other hand, $G$ decreases as the melt pools become larger with increasing $P$ for a constant $V$. On average, $G$ is minimum ($0.6 \times 10^7$ K m$^{-1}$) for the largest melt pool (Case 5) and maximum ($1.5 \times 10^7$ K m$^{-1}$) for the smallest melt pool (Case 1). As expected, $R$ is minimum (0.05 m s$^{-1}$) for the cases with smallest $V$ and maximum (0.2 m s$^{-1}$) for the cases with largest $V$.
\subsection{Phase-field simulations}
\begin{figure*}[ht]
\begin{center}
\includegraphics[scale=0.65]{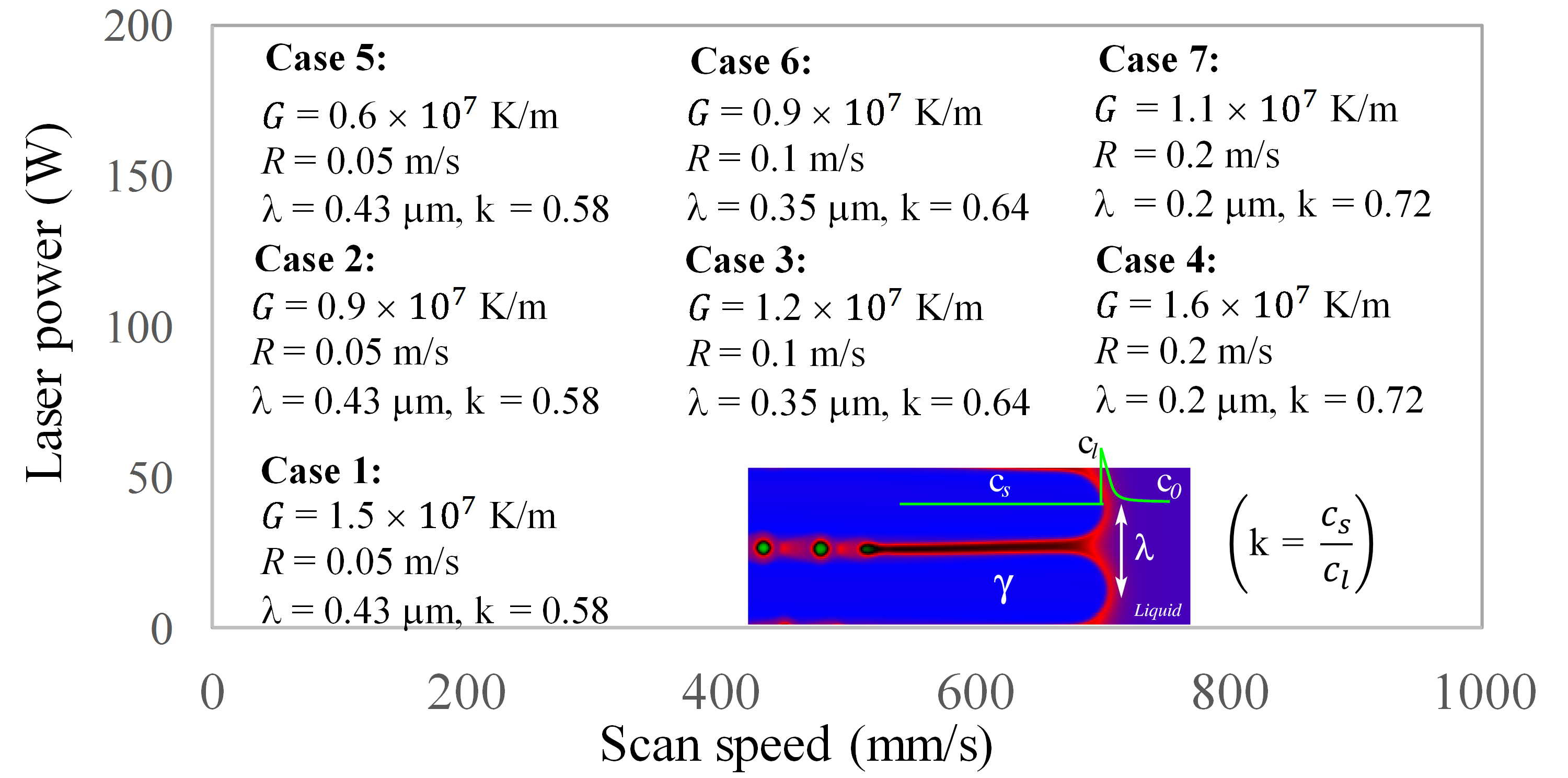}
\caption{A snapshot of typical cellular microstructure at steady state is presented for 195 W and 800 mm s$^{-1}$ (bottom right). Two main features of the $\gamma$ cells are reported, the primary spacing $\lambda$ and the Nb distribution coefficient $k$, as functions of average temperature gradient $G$ and the solidification rate $R$. $\lambda$ decreases by almost 20\% as $R$ increases by a factor of $2$. Cell core composition $c_s$ (= 3.7\%) remains constant as the diffusivity of solid is ignored. Since Nb is rejected as the cells grow, there is a spike in Nb composition $c_l$ (= 5.1\%), which decays exponentially to reach the nominal composition $c_0$ (= 4.0\%) in the far-field liquid. The distribution of Nb across the cell-liquid interface is thus represented by the non-equilibrium distribution coefficient $k = c_s/c_l$. Similar to $\lambda$, the effect of $P$ is negligible on $k$ which increases with increasing $V$. $k$ deviates by 50\% from the local equilibrium ($k = 0.48$) for the cases with highest $V$.}\label{fig_phase-field}
\vspace{-8mm}
\end{center}
\end{figure*}

With the preceding calculations of the melt pool parameters, it is now possible to model the solidification process in the melt pool. We use a binary alloy phase-field model~\cite{Echebarria2004} to simulate $\gamma$-Ni cells and microsegregation of Nb (nominal mass fraction of 4\%) as a binary approximation of an IN625 microstructure. The time-dependent order parameter ($\phi$) and concentration ($c$) model equations of motion are solved on a uniform mesh (480$\times$5000), using the finite volume method, explicit time stepping scheme and zero-flux boundary conditions. Since the solid-liquid interface equilibrium does not hold during AM solidification~\cite{supriyo2017}, the present simulations are conducted for vanishing anti-trapping solute flux term (second term inside bracket in Eq.~(5) in~\cite{Trevor2017,supriyo2017}) to model realistic solute redistribution across the solid-liquid interfaces. The original purpose of adding this extra solute flux term to the standard Fickian diffusion flux in the concentration equation was meant to enforce local equilibrium during low velocity casting solidification simulations~\cite{Echebarria2004}. The solid-liquid interfacial energy $\sigma$ = 0.3 J m$^{-2}$~\cite{Hoyt2003}, the solid-liquid cubic anisotropy parameter $\epsilon_4$ = 0.02~\cite{Hoyt2003} and the solid-liquid interface thickness $W_0$ = \SI{5}{\nano\meter} are approximated for a Ni-Nb alloy. For the model equations and other numerical and thermophysical parameters, refer to ~\cite{Trevor2017,supriyo2017}. Each simulation begins with a thin layer of solid at the bottom of the simulation box, accompanied with a small, random amplitude perturbation in $\phi$ at the initial solid-liquid interface, from which stable perturbations grow with time and break into steady state $\gamma$-Ni cells (Fig.~\ref{fig_phase-field}).

The microstructures that form during laser resolidification are governed by the combinations of $G$ and $R$. $G/R$ determines the nature of the solidification front (planar, cellular or dendritic) and $GR$ (cooling rate) controls the size of the solidification structure~\cite{kurzbook,Rappazbook}. In our finite element calculations, $G/R$ varies between 80 K~s~mm$^{-2}$ and 300 K~s~mm$^{-2}$ and $GR$ varies between $3\times10^6$ K s$^{-1}$ and $3\times10^7$ K s$^{-1}$. In this $G-R$ space, our simulations predict columnar dendritic morphology of the Ni-rich $\gamma$ matrix phase. The average distance between the $\gamma$ cells remains constant at steady state (Fig.~\ref{fig_phase-field}), which is the primary dendrite arm spacing $\lambda$. We use the Fourier spectrum analysis~\cite{Greenwood2004} of the solid-liquid interface of the cellular microstructures to estimate $\lambda$. In our simulations, $\lambda$ decreases with increasing $GR$ and ranges between \SI{0.20}{\micro\meter} and \SI{0.43}{\micro\meter}, which can be approximated following Hunt~\cite{Hunt1979book}: $\lambda$(\SI{}{\micro\meter}) = $ A (k \Gamma \Delta T_0 D_l)^{0.25} G^{-0.5}R^{-0.25}$, where $k$ is the partition coefficient, $\Gamma$ the Gibbs Thomson coefficient, $\Delta T_0$ the equilibrium freezing range, $D_l$ the diffusivity of the liquid, and the coefficient $A$ is calculated between 0.13 to 0.17. Such extremely dense cellular microstructures improve the yield, tensile and fatigue strengths of the as-deposited material and reduce melt flow effects~\cite{Tan2011}. $\lambda$ decreases significantly with increasing $V$, while $\lambda$ does not seem to be affected by increasing $P$ in our simulations. 

Niobium is the most important alloying element in IN625. Niobium segregates most severely from the solid cell core ($c_s$) to the liquid ($c_l$) due to the smallest equilibrium partition coefficient ($k = c_s/c_l = 0.48$) among all elements in IN625, and determines the solidification pathways and the precipitation of strengthening phases in the solid state~\cite{Liu2013}. Our simulations show strong microsegregation of Nb in between the cells, illustrating the presence of non-equilibrium Nb redistribution during single line laser resolidification. As a result, $k$ becomes a function of the solid-liquid interface growth rate $R$ and ranges between $0.58 \leq k \leq 0.72$ (Fig.~\ref{fig_phase-field}). Note that the limit $k = 1$ is reportedly reached only at $R$ $\geq$ 9 m~s$^{-1}$~\cite{Kurz1994}. In contrast to $\lambda$, $k$ increases significantly with increasing $R$ (or $V$), while $k$ does not seem to be affected with increasing $P$. Such a response of the solidifying Ni-Nb melt pool boundary due to the laser treatment can be characterized using the Aziz~\cite{Aziz1982} solute-trapping function given by $k = (1 + R/V_d)/(0.48 + R/V_d)$, where  $V_d$ = 0.23 m s$^{-1}$ is the diffusion velocity of Nb during laser resolidification in IN625. The Nb-rich liquid droplets (Fig.~\ref{fig_phase-field}) that shed from the root region of the dendrites subsequently freeze, leading to regions of highest microsegregation which could transform to intermetallic phases in the solid state~\cite{Zhang2017_nist}.
 
\section{Discussion and Conclusions}
In this work, we used experimental measurements of the melt pool dimensions ($l$, $w$ and $h$) to validate heat conduction based finite element simulations for the $P-V$ map of single track laser treatments. The size of the melt pool increases with increasing $P$ and decreasing $V$. Our simulations showed reasonable match of $l$, $w$, $h$, and $l/w$ with experiments except when keyhole melting is observed experimentally. The $l/w$ ratio calculated in this case is $3.2$, which is close to the value when the melt pool dimensions become a function of the surface tension, recoil pressure and other hydrodynamic effects~\cite{King2015,Childs2005} which affect the shape and stability of the keyhole. The laser absorptivity also becomes an increasing function of the beam power, which we model as a constant ($\eta$ = 0.5) for simplification reasons. Note that keyhole solidification often results in porosity, leading to poor mechanical behavior of the parts and is therefore not recommended for AM processing~\cite{Montgomery2015,Rai2007}. The finite element simulations in Ref.~\cite{Jon2017} are particularly interesting in that they characterize the melt pool dynamics on its dimensions using a high power laser beam ($P$ = 500 W) and it was observed that the simulated melt pool dimensions increasingly deviated from the experimental measurements with increasing $P$. For a system with low to moderate power of the laser beam and very small melt pool depth ($h$~$<$~\SI{50}{\micro\meter}), our simulations may be appropriate, as discussed in~\cite{Guo2000,Suresh2016}.

As demonstrated by multiphysics modeling of the laser-powder-substrate interaction in Ref.~\cite{King2015}, the vaporization-induced recoil force on the melt pool typically produces a ``hollow'' under the laser beam that extends through the powder layer into the substrate. This is even true in the conduction regime where additive manufacturing is typically carried out. Such behavior cannot be readily incorporated into a finite element model, so all such models must be seen as severe approximations, and the best way we know to validate such a model is to test if the predicted melt geometry is correct. That is the approach taken here. If the steady-state melt pool length and cross-sectional size and shape are correct, then the resulting thermal behavior as a function of time and position should be adequate for microstructure evolution studies. One drawback to this approach is that it is only valid for one material system at a time, IN625 in this case.  Also, although the current finite element model should be adequate for the results reported here, it should be mentioned that although the melt pool lengths agree well with the \textit{in situ} melt-pool measurements, the shapes of the melt pools are not perfect. Decreasing the absorptivity and slightly increasing the width of the laser beam would likely improve the agreement, and work in this direction is in progress along with \textit{in situ} measurements of the absorptivity~\cite{Matthews2017}.

The melt pool cooling rates estimated from both experiments and simulations are on the same order of magnitude $\approx$ $10^6$ K s$^{-1}$. The microstructural features that form in these cooling rates in our simulations change primarily as a function of $R$, signifying that $R$ is a more important parameter than $G$ during AM. Previous studies have shown that $G/R$~$<$~1 K~s~mm$^{-2}$ resulted in an equiaxed mode of dendritic solidification, 1 K~s~mm$^{-2}$ $<$ $G/R$~$<$~80~K~s~mm$^{-2}$ resulted in a mixed equiaxed and columnar solidification, and $G/R$~$>$ 80~K~s~mm$^{-2}$ resulted in columnar solidification in Inconel alloys~\cite{Raghavan2016}. In our simulations, $G/R$ ranges between 80~K~s~mm$^{-2}$ s and 300~K~s~mm$^{-2}$ and produces columnar dendritic morphologies. The $G-R$ solidification map is determined by various combinations of $P$ and $V$ and therefore can be used, for example, to engineer the appropriate grain morphology and grain size for different sections of the AM parts. The simulations of melt pool solidification in Ref.~\cite{Ranadip2017} are particularly interesting where an initial solid-liquid interface, similar to the resolidification boundary in our Fig.~\ref{fig_microstructure}, resulted in local nucleation and growth of multiple dendrites with a misorientation with the vertical direction between 0$^\circ$ to 20$^\circ$ depending on the scan velocity, thermal gradient, interface roughness and initial seed position.

Unlike finite element simulations, the microstructure evolution for the calculated $G-R$ solidification map does not depend much on melt pool dynamics. This is precisely due to the extremely fine cellular structures, which provide significant resistance to fluid flow following an exponential increase of the damping effect in the semisolid mushy region~\cite{Tan2011}. Phase-field simulations performed on Ni-Nb alloys showed that the solute redistribution across dendrites remained similar when simulations were conducted with and without convection~\cite{Lee2010}. The effects of convection on the spacing between primary dendrite arms is not as pronounced as compared to the spacing between side arms~\cite{Lee2010}, which we do not observe in our simulations. Therefore, melt pool solidification is modeled with reasonable approximations for the same parameter combinations in both experiments and simulations. Further research is required for more accurate characterization of the melt pool dimensions and multicomponent solidification microstructures~\cite{Ghosh2017_eutectic} with and without melt pool dynamics in order to improve the predictive capabilities for the quality of the as-built parts, although it will be computationally intensive.

\bibliography{papers} 
\bibliographystyle{spmpsci}      

\end{document}